\documentstyle[12pt]{article}
\begin{document}
\thispagestyle{empty}
\begin{flushleft}
\end{flushleft}
\thispagestyle{empty}
\begin{flushleft}
{\Large \bf 
DROPLETS IN GENERAL RELATIVITY: EXACT SELF-CONSISTENT 
SOLUTIONS TO THE INTERACTING SCALAR AND ELECTROMAGNETIC FIELD EQUATIONS}
\end{flushleft} 
\hskip 2 cm
\vbox{ 
\hbox{\bf Yu.\ P.\ RYBAKOV and G.\ N.\ SHIKIN} 
\hbox{\it Department of Theoretical Physics} 
\hbox{\it Russian Peoples' Friendship University} 
\hbox{\it 6, Miklukho-Maklay str., 117198 Moscow, Russia} 
\hbox{\it e-mail: rybakov@udn.msk.su}
\vskip 3mm 
\hbox{\bf B.\ SAHA} 
\hbox{\it Laboratory of Theoretical Physics} 
\hbox{\it Joint Institute for Nuclear Research, Dubna} 
\hbox{\it 141980 Dubna, Moscow region, Russia} 
\hbox{\it e-mail:  saha@thsun1.jinr.dubna.su}} 
\vskip 1.5cm
\noindent
It is shown, for the self-consistent system of scalar, electro-magnetic 
and gravitational fields in general relativity, that the equations of 
motion admit a special kind of solutions with spherical or cylindrical 
symmetry. For these solutions, the physical fields vanish and the space-time 
is flat outside of the critical sphere or cylinder. Therefore, the mass and 
the electric charge of these configurations are zero. The principal 
difference between droplet-like solutions with spherical symmetry and 
those with cylindrical one has been  established.  In  the first  case 
there  exists  a  possibility  of  continuous  transformation of 
droplet-like configuration into the solitonian one, when for  the second 
case there is no such a possibility.

\newpage
\section{Introduction} 
\setcounter{equation}{0}
A specific type of solutions to the nonlinear field equations in flat 
space-time were obtained in a series of interesting articles [1--4]. These 
solutions are known as droplet-like solutions or simply droplets. 
Distinguishable  property of these solutions is the availability of some 
sharp boundary, defining the space domain, in which the material 
field happens to be located i.e. the field is zero beyond this area. As 
was found the solutions mentioned exist in field theory with specific 
interactions that can be considered as effective, generated by initial 
interactions of the unknown origin.

The aim of this paper is to consider self-consistent system to obtain 
droplet-like solutions in the framework of General Theory of Relativity. 
We show that in case of interacting electromagnetic and gravitation field 
system with specific type of interactions there exist droplet-like 
solutions having zero electric charge and mass. It is noteworthy to 
underline that the effective potentials, raised in this case, possess 
confining property i.e. create a strong repulsion in some configurational 
space.

   We study the existence of regular static solutions, possessing 
spherical and/or cylindrical symmetry and sharp boundary, to  the 
equations  of  scalar  nonlinear   electrodynamics   in   General 
Relativity. Excluding the scalar field we  obtain  the  effective 
Lagrangian of nonlinear electrodynamics. Contrary to  the  widely 
known soliton-like solutions, with field  functions  and  energy 
density asymptotically tending to zero at spatial infinity,  the 
solutions in question vanish at a finite distance from the center 
of the system (in the case of spherical  symmetry)  or  from  the 
axis (in the case of cylindrical symmetry).  Thus, there  exists 
the sphere or cylinder with critical radius $r_0$, outside of which 
the fields disappear. Therefore the field configurations have the 
droplet-like structure [1,\ 5,\ 6].  As  is  known, there  do  not  exist 
regular   static   spherically   or    cylindrically    symmetric 
configurations within the framework of gauge invariant  nonlinear 
electrodynamics [7]. That is why we consider the generalized model, 
with  the  Lagrangian  explicitly  containing 4-potential ${\cal A}_\mu$,  
\quad $\mu\,=\, 0,\ 1,\ 2,\ 3,$ thus breaking the gauge invariance inside 
the critical sphere or cylinder. The corresponding terms  appear  due  to  
the interaction between the electromagnetic and scalar  fields.  This 
interaction being negligible at large distances, the  Maxwellian 
structure of the electromagnetic Lagrangian (and  therefore  the gauge 
invariance) is reinstated for $r\, \gg \, r_0 $.
\section{Fundamental Equations}  
\setcounter{equation}{0}

We choose the Lagrangian in the form [7]
\begin{equation}
{\cal L}\,=\, {\cal R}/ 2\,\kappa \,-\, (1/ 16\,\pi)\, \bigl[{\cal 
F}_{\alpha \beta}\,{\cal F}^{\alpha \beta}\,+\, 2\, \varphi_{, \alpha}\, 
{\varphi}^{, \alpha}\,\Psi(I)\,\bigr],
\end{equation}
where $\kappa\,=\, 8\,\pi\,G$ is the Einstein's  
gravitational constant and the function $\Psi(I)$ of the invariant $I\,=\, 
{\cal A}_\mu\,{\cal A}^\mu$ characterizes the interaction between the 
scalar $\varphi$ and electromagnetic ${\cal A}_\mu$ fields.  In the sequel 
there will not appear any restrictions  of  the function $\Psi(I)$, thus 
the Lagrangian (2.1) defines  the class  of models parameterized by 
$\Psi(I)$.  

The field equations corresponding to the Lagrangian (2.1) read 
\begin{equation}
{\cal G}_{\mu}^{\nu}\,=\,-\,\kappa\,T_{\mu}^{\nu},
\end{equation}
\begin{equation}
\partial _\alpha\, \bigl(\sqrt{-g}\,g^{\alpha \beta}\, \varphi_{, \beta} 
\, \Psi \bigr)\,=\,0, 
\end{equation}
\begin{equation}
(1/\sqrt{-g})\, \partial_\beta \bigl(\sqrt{-g}\, {\cal F}^{\alpha \beta} 
\bigr)\,+\, \bigl(\varphi_{, \beta}\,{\varphi}^{, \beta}\bigr)\, \Psi_I \, 
{\cal A}^\alpha\,=\,0, 
\end{equation}  
where $\Psi_I\,=\, d\Psi/dI$ and ${\cal G}_{\mu}^{\nu} \,=\, {\cal 
R}_{\mu}^{\nu}\, -\, \delta_{\mu}^{\nu}\, {\cal R} /2 $ is the Einstein 
tensor. One can write the energy-momentum tensor of the interacting matter 
fields in the form:  
$$
T_{\mu}^{\nu}\,=\,(1/4 \pi)\,\bigl[\varphi_{, \mu}\, \varphi^{, \nu}\, 
\Psi(I)\,-\,{\cal F}_{\mu \alpha}\,{\cal F}^{\nu \alpha}\,+\,
\varphi_{, \alpha}\, \varphi^{, \alpha}\,\Psi_I\,{\cal A}_\mu\,{\cal 
A}^\mu \bigr]\,+  $$
\begin{equation}
 +\, \delta_{\mu}^{\nu}
\bigl[(1/ 16\,\pi)\, \bigl({\cal F}_{\alpha 
\beta}\,{\cal F}^{\alpha \beta}\,+\, 2\, \varphi_{, \alpha}\, 
{\varphi}^{, \alpha}\,\Psi(I)\,\bigr)\bigr]. \end{equation}

\section{Configurations with spherical symmetry}
\setcounter{equation}{0}
   Searching for the static  spherically-symmetric  solutions  to 
the system of equations (2.2)\,-\,(2.4) , we consider the metric  in  the 
form [8]:
\begin{equation}
ds^2\,=\,e^{2\gamma}\,dt^2\,-\,e^{2\alpha}\,d\xi^2\,-\, 
e^{2\beta}\bigl[d\theta^2\,+\,\mbox{sin}^2\theta\ d\phi^2 \bigr]. 
\end{equation}
Let us now formulate the requirements to be fulfilled by particle-like 
solutions (PLS). These are [9]

(a) Stationarity [applied to the metric (3.1)] i.e.
$$ \alpha\,=\,\alpha(\xi), \quad \beta\,=\,\beta(\xi), \quad \gamma\,=\,
\gamma(\xi);$$

(b) regularity of the metric and the matter fields in the whole 
space-time;

(c) asymptotically Schwarzschild metric and corresponding behavior of the 
field functions.

In view of requirement (a) it is convenient to choose the harmonic $\xi$
coordinate ($\Box \xi\,=\,0$) in (3.1) to satisfy the subsidiary condition 
[10]:  \begin{equation} \alpha\,=\,2\,\beta\,+\,\gamma.  \end{equation} 
The corresponding coordinate in flat space-time is just $\xi\,=\,1/r.$

With the constraint (3.2) the system of Einstein equations (2.2) reads:  
\begin{equation}
e^{-2 \alpha}\,(2 \beta^{\prime \prime}\,-\, U)\,-\, e^{-2 \beta}\,=\, 
-\kappa\,T_{0}^{0},
\end{equation}
\begin{equation}
e^{-2 \alpha}\, U \,-\, e^{-2 \beta}\,=\, 
-\kappa\,T_{1}^{1},
\end{equation}
\begin{equation}
e^{-2 \alpha}\,(\beta^{\prime \prime}\,+\,\gamma^{\prime \prime}\,-\, 
U)\,=\, -\kappa\,T_{2}^{2}\,=\,-\kappa\,T_{3}^{3},
\end{equation}
where $U\,=\, \beta^{\prime 2}\,+\, 2\,\beta^{\prime}\,\gamma^{\prime}$,
and $\prime\,\equiv\, d/d\xi.$  

Note that the field functions, as well as 
the components of  the metric tensor depend  on the single 
spatial variable $\xi$.  Assuming the electromagnetic field to be 
determined by the time  component ${\cal A}_0 \,=\,{\cal A}(\xi)$ of the  
4-potential one finds the  unique non-trivial component of the field  
tensor  ${\cal F}_{10} \,=\, {\cal A}^{\prime},$ and  the invariant $I$ 
reduces to $I\,=\, e^{-2\gamma}\,{\cal A}^2(\xi).$  

One can write the  non-zero  components  of  the  energy-momentum 
tensor (2.5) as follows:
\begin{equation}
T_{0}^{0}\,=\,(1/8\pi)\,e^{-2\alpha}\,\bigl[{\cal A}^{\prime 
2}\,e^{-2\gamma} \,+\, \varphi^{\prime 2}\bigl(\Psi\,-\,2\,{\cal A}^2\, 
e^{-2\gamma}\,\Psi_I\,\bigr)\bigr],
\end{equation} 
\begin{equation}
T_{1}^{1}\,=\,-\,T_{2}^{2}\,=\,-\,T_{3}^{3}\,=
\,(1/8\pi)\,e^{-2\alpha}\,\bigl[{\cal A}^{\prime 
2}\,e^{-2\gamma} \,+\, \varphi^{\prime 2}\,\Psi\bigr].
\end{equation} 
Adding together the equations (3.4) and (3.5) and using the property
$T_{1}^{1}\,+\,T_{2}^{2}\,=\, 0$, one obtains the differential equation
$$\beta^{\prime \prime}\,+\,\gamma^{\prime 
\prime}\,-\,e^{2(\beta\,+\,\gamma)}\,=\,0,$$                       
with the solution [11]
\begin{equation}
e^{-(\beta\,+\,\gamma)}\,=\,{\cal 
S}(k,\,\xi)\,=\,\left\{\begin{array}{ccc} k^{-1}\,\mbox{sh}\,k\xi,& 
k\,>\,0, \\ \xi,& k\,=\,0, \\ k^{-1}\, \mbox{sin}\,k\xi,& k\,<\,0, 
\end{array}\right .
\end{equation}
depending on the 
constant $k$. Notice  that  another  constant  of integration is trivial, 
so that $\xi \,=\, 0$ corresponds to the  spatial infinity, where 
$e^\gamma \,=\,1$ and $e^\beta \,=\,\infty$.  Without loss of generality 
one can choose $\xi \,>\, 0.$  

The scalar field equation (2.3) has the evident solution 
\begin{equation}
\varphi^{\prime}\,=\,C\,P(I),
\end{equation} 
where $P(I)\,=\, 1/\Psi(I)$ and $C$ is  the  
integration constant.  Putting (3.9) into (2.4) one gets the equation for 
the electromagnetic field 
\begin{equation}
\bigl(e^{-2\gamma}\,{\cal A}^{\prime}\bigr)^{\prime}\,-\, C^2\, P_I\, 
e^{-2\gamma}\,{\cal A}\,=\,0,
\end{equation} 
where the second term could  be naturally interpreted  as  the 
induced nonlinearity.  
In view of (3.9) one rewrites the Einstein equation 
(3.4) and the result of adding the equations (3.3) and (3.4) as follows :  
\begin{equation}
\gamma^{\prime 2}\,=\,-G\,\bigl(C^2\,P\,-\,{\cal A}^{\prime 
2}\,e^{-2\gamma} \bigr)\,+\,K, \qquad  K\,=\, k^2\,\mbox{sign}k, 
\end{equation}
\begin{equation}
\gamma^{\prime \prime}\,=\,G\,e^{-2\gamma}\,\bigl({\cal A}^{\prime 2}\, 
+\,C^2\,{\cal A}^2\,P_I \bigr).
\end{equation} 
One can easily check that the 
equation (3.11) is the first integral of the equations (3.10) and (3.12).  

Eliminating the term $(P_I\,{\cal A})$ between (3.10) and (3.12) one gets 
the equation:  
\begin{equation}
\gamma^{\prime \prime}\,=\,G\,\bigl({\cal A}\,{\cal 
A}^{\prime}\,e^{-2\gamma}\bigr)^{\prime}, \end{equation} 
with the evident first integral: 
\begin{equation}
\gamma^{\prime}\,=\,G\,{\cal A}\,{\cal 
A}^{\prime}\,e^{-2\gamma}\,+\,C_1, \quad C_1\,=\,\mbox{const.} 
   \end{equation} Let us consider the simple case $C_1 \,=\,0.$ Then from 
(3.14) we get \begin{equation} e^{2\gamma}\,=\,G\,{\cal A}^2\,+\,H,\qquad 
H\,=\,\mbox{const.}  \end{equation} Substituting $\gamma^{\prime}$ and 
$e^{2\gamma}$ from (3.14) and (3.15) into (3.11), we find for ${\cal 
A}(\xi)$ the differential equation:  
\begin{equation} {\cal A}^{\prime 
2}\,(G\,{\cal A}^2\,+\,H)^{-2}\,=\, (G\,C^2\,P\,-\,K)/\,G\,H, 
\end{equation}  
which can be solved by quadrature:  
\begin{equation}
\int\limits_{}^{}\frac{d {\cal A}}{(G\,{\cal A}^2\,+\,H)\, 
\sqrt{G\,C^2\,P\,-\,K}}\,=\,\pm\,(1/\sqrt{G\,H})\, (\xi\,-\,\xi_0), \quad
\xi_0\,=\,\mbox{const.}
\end{equation} 
It is clear that the 
configuration obtained has a center if and only if $e^{\beta} \,=\, 0$ at 
some $\xi\,=\, \xi_c.$  One  can  show [10] that the conditions for the 
center $\xi_c \,=\, \infty$ to be regular imply $K\,=\,0$ and  the 
following behavior of the field quantities in the vicinity of the point 
$\xi_c\,=\,\infty$:
\begin{equation}
\gamma^{\prime}\,=\,O\,(\xi^{-2}), \quad {\cal A}^{\prime} \to 
{\cal A}_c\,\ne\,\infty, \quad {\cal A}^{\prime}\to 0, 
\end{equation} 
\begin{equation}
\xi^4\,P(I) \to 0, \quad \mid \xi^4\,I\,P_I \mid\,<\,\infty.
\end{equation}
In view of (3.18) we deduce from (3.14) that $C_1 \,=\,0$ in accordance 
with the earlier supposition.  

Now we can write the boundary conditions on 
the surface  of the critical sphere $\xi\,=\,\xi_0$:
\begin{equation}
T_{\mu}^{\nu}\,=\,{\cal A}\,=\,{\cal A}^{\prime}\,=\,0, \quad e^{\gamma}\, 
=\,1,\quad e^{\beta}\,=\, 1/\xi_0\,>\,0.
\end{equation}
Due to (3.20) and (3.15) we infer that $H=1.$  

Let us now choose the function $P(I)$ as follows [12]:  
\begin{equation}
P(I)\,=\,J^{(1\,-\,2/\sigma)}\,\bigl[(1\,-\,J)^{1/\sigma}\,-\, 
J^{1/\sigma} \bigr]^2\,(1\,-\,J),
\end{equation}  
where $J\,=\, G\,I;\quad \sigma\,=\, 2n\,+\,1;\quad  n=1,2,3\cdots$  
Then on account of $K\,=\, 0$ and $H\,=\,1$ we get from (3.17)  the 
following expression for ${\cal A}(\xi)$:  
\begin{equation}
{\cal A}(\xi)\,=\,(1/\sqrt{G})\,\bigl[1\,-\, 
\mbox{exp}\,\bigl(-\frac{2\,C\,\sqrt{G}}{\sigma}\,(\xi\,-\,\xi_0)\bigr) 
\bigr]^{\sigma/2}.
\end{equation}  

As one can see from (3.22), the 
conditions (3.18) and (3.19) for the center to be regular and the 
matching conditions (3.20) on  the surface of the critical sphere are 
fulfilled if $\sigma\,>\,2.$ It is also obvious from (3.22) that for $\xi 
< \xi_0$ the value of square bracket turns to be negative one and ${\cal 
A}(\xi)$ becomes imaginary since $\sigma$ is an odd number. Since we are 
interested in the real ${\cal A}(\xi)$ only, without loss of generality 
we may assume the value of ${\cal A}(\xi)$ be zero for $\xi \le \xi_0$. 

Recalling that $J\,=\,G\,{\cal A}^2 /\,(G\,{\cal A}^2 \,+\,1)$, we get 
from (3.22) that $J(\infty)\,=\, 1/2$ and $J(\xi_0)\,=\,0$, thus implying: 
\begin{equation}
P(I)\mid_{\xi\,=\,\infty} \,=\,P(I)\mid_{\xi\,=\,\xi_0}\,=\,0.
\end{equation}
It means that at $\xi\,=\,\xi_c\,=\,\infty$ and $\xi\,=\,\xi_0$, the 
interaction function $\Psi(I)\,=\,1/P(I)$ is singular.  It  turns out 
nevertheless that  the energy density $T_{0}^{0}$  is regular at these 
points due to the fact that it contains $\Psi(I)$ as a multiplier in the 
form:  
\begin{equation}
e^{-2\alpha}\,\varphi^{\prime 2}\,\Psi\,=\,C^2\,e^{-2 \alpha}\,P(I),
\end{equation}
which tends to zero as $\xi\,\to\,\xi_c$ or $\xi\,\to\,\xi_0.$  

As follows from (3.22), for the limiting case $\xi_0\,=\,0$, when the 
critical sphere goes to the spatial infinity and the solution in question 
is defined at $0\,\le\,\xi\,\le\,\infty$, it appears that  at spatial 
infinity $(\xi\,=\,0)\quad {\cal A}\,=\,0$ and $P(I)\,=\,0$. In this case 
we obtain the usual soliton-like configuration not possessing any sharp 
boundary.  

Note that at spatial infinity $(\xi\,=\,0)$ one can compare the metric 
found with the Schwarzschild one and the electrical field with the Coulomb 
one, thus determining the total mass $m$ and the charge $q$ of the system:  
$$G\,m\,=\, -\,\gamma^{\prime}(0),\qquad q\,=\,-\,{\cal A}^{\prime}(0).$$ 
Taking into account that $e^{2\gamma}\,=\,G\,{\cal A}^2 \,+\,1$, one can 
find through the use of (3.22) that for $\xi_0\,=\,0,\quad {\cal 
A}^{\prime}(0)\,=\,-q\,=\,0$ and $\gamma^{\prime}(0)\,=\,-G\,m\,=\,0.$  
Therefore, the total energy of the soliton-like system,defined as the sum 
of  the material fields energy and that of the gravitational field, 
vanishes.  If now one chooses the integration constant $\xi_0\,>\,0$,  
then the field configuration with the sharp boundary (droplet) appears. In 
this case for $\xi\,\le\,\xi_0$  one obtains ${\cal A}(\xi)\,=\,0$ and  
$e^{2\gamma}\,=\,1$, i.e.  outside of the droplet gravitational and 
electromagnetic fields disappear, that implies the vanishing  of the total 
mass  and  the charge of  the system.  This unusual property makes the 
droplet-like object poorly visible  for the outer observer.  
   
Let us now 
calculate the matter field energy density: 
\begin{equation}
T_{0}^{0}\,=\,(C^2/\,8\,\pi)\,e^{-2\alpha}\,\bigl[\,P\,(1\,+\,e^{2\gamma} 
)\,+\,2\,I\,P_I(I)\,\bigr].
\end{equation} 
To this end we  substitute into (3.25)  the expression 
for $e^{-2\alpha}\,=\,\xi^4\,(G\,A^2 \,+\,1),\,\, P(I)$ and ${\cal 
A}(\xi)$ using (3.21) and (3.22).  It should be emphasized that the field 
energy is localized in a small region $(\xi_0 \,\le\,\xi\,<\,\infty):$
\begin{equation}
T_{0}^{0}(\xi)\mid_{\xi\to \infty}\,\to \,0, \qquad
T_{0}^{0}(\xi)\mid_{\xi\to \xi_0}\,\to \,0, 
\end{equation} 
namely, inside the critical sphere with the 
radius 
$$ R \,=\, \int\limits_{0}^{\infty}\,d\xi\,e^{\alpha(\xi)}
\,=\, \int\limits_{0}^{\infty}\,d\xi\,/\xi^2\Bigl\{\bigl[1 - e^
{-2C\sqrt{G}(\xi-\xi_0)/\sigma}\bigr]^\sigma +1\Bigr\}^{(1/2)}
\,< \infty .$$  
One can readily derive from (3.25) the energy $E_f$ of the matter fields:  
\begin{equation}
E_f\,=\,\int\limits_{}^{}\,d^3 x\,\sqrt{-^3\,g}\,T_{0}^{0}\,=\,
(C/2)\,\int\limits_{0}^{1/\sqrt{G}}\,d{\cal A}\,e^{-3\gamma}\, 
\bigl[\sqrt{P}(1\,+\,e^{2\gamma})\,+\,4\,I\,(\sqrt{P})_I\,\bigr], 
\end{equation} 
Using (3.21) and the expressions 
$$ e^{2\gamma}\,=\,1/(1\,-\,J), \qquad {\cal A}\,=\,\sqrt{G\,P\,(1\,-\, 
J)}, $$ 
one finds from (3.27) after integrating by parts, 
$$ E_f\,=\,(C/4\sqrt{G})\,\int\limits_{0}^{1/2} 
\bigl[(2\,-\,J)\sqrt{J\,P}/(1\,-\,J)\,+\,2\,P_J\,\sqrt{J^3/P}\bigr]\,dJ\,=$$
$$ =\,(C/4\sqrt{G})\,\int\limits_{0}^{1/2}\,[(5\,J\,-\,4)\,\sqrt{J\,P}/ 
(1\,-\,J)]\,dJ\,<\,0.$$ 
Knowing that the total energy of the droplet-like  
object is zero this inequality implies the positivity of its gravitational 
energy. Thus the droplet-like configuration of the fields obtained is 
totally regular with zero total energy (including the energy of 
proper gravitational field) and null electric charge and remains 
unobservable to one located outside the sphere with radius $R$ [12,\ 13]. 

In order to clarify the fact that the role of the gravitational 
field in forming the droplet-like configuration  is not decisive it is 
worthwhile to compare the solution obtained with that in the flat 
space-time, described by the interval 
$$ ds^2\,=\,dt^2\,-\, dr^2\,-\,r^2\,[d\theta^2\,+\, \mbox{sin}^2\theta\, 
d\phi^2].$$  
In the latter case the equation (2.3) 
admits the solution 
\begin{equation}
\varphi^{\prime}(r)\,=\,C\,P(r)/r^2.
\end{equation}
Substituting (3.28) into (2.4), 
one finds that the equation for the electromagnetic field can be solved by 
quadrature:  
\begin{equation}
\int\limits_{}^{} d{\cal A}/\sqrt{P}\,=\,\pm\,C\,\bigl(\frac{1}{r}\,-\, 
\frac{1}{r_0}\bigr), \quad r_0\,=\,\mbox{const.}
\end{equation}   
Note that the droplet-like configuration ${\cal A}(r)$ will be similar to 
(3.22) if one chooses the function $P(I)$ more simple than (3.21):  
\begin{equation}
P(I)\,=\,J^{1\,-\,2/\sigma}\,(1\,-\,J^{1/\sigma})^2, \qquad 
J\,=\,\lambda\,I,\end{equation}  
where $\lambda\,=\,$ const;\quad $\sigma\,=\,2\,n\,+\,1;\quad 
n\,=\,1,2,3,\cdots$. Then substituting (3.30) into (3.29) one gets the 
solution 
\begin{equation}
{\cal 
A}(r)\,=\,(1/\sqrt{\lambda})\,\bigl[1\,-\,\mbox{exp}\bigl(-\frac{2\,C\,
\lambda}{\sigma}\,(\frac{1}{r}\,-\,\frac{1}{r_0})\bigr)\bigr]^{\sigma/2}. 
\end{equation} 
One can see from (3.31) that ${\cal A}(r)\,=\,0$ as $r\,\ge\,r_0$, i.e. 
the charge of the flat space-time droplet configuration also vanishes.  
For this solution the regularity conditions at the center $r\,=\,0$ and  
on the surface of the critical sphere $r\,=\,r_0$  are evidently 
fulfilled.  It similarly appears that for $r\,=\,\infty$  one finds the 
usual soliton-like structure with field vanishing as $r\,\to \infty$.  
The field energy $E_f$ is defined as follows:  
\begin{equation}
E_f\,=\,C\,\int\limits_{{\cal A}(r_0)}^{{\cal A}(0)}\,d{\cal A}\, 
(\sqrt{P} \,+\,I\,P_I/\sqrt{P})\,=\,C\sqrt{P\,I}\mid_{{\cal 
A}(r_0)}^{{\cal A}(0)}.  
\end{equation}  
Inspecting that $P\,I\,=\,0$ both at $r\,=\,0$ and $r\,=\,r_0$, we arrive 
through (3.32) at $E_f \,=\,0.$  

Thus in the flat space-time as well as 
for the self-gravitating system, the total energy and charge of the 
droplet-like configuration vanish.  

\section{Configurations with cylindrical symmetry}
\setcounter{equation}{0}

Let us now search for static cylindrically-symmetric solutions to 
the equations (2.2)-(2.4). In this case the metric can  be chosen as 
follows [14,\ 15]:  
\begin{equation}
ds^2\,=\,e^{2\gamma}\,dt^2\,-\,e^{2\alpha}\,dx^2\,-\, 
e^{2\beta}\,d\phi^2\,-\,e^{2\mu}\,dz^2. 
\end{equation}

The requirements to be fulfilled by soliton-like 
solutions (PLS) in this case are [16]

(a) Stationarity [applied to the metric (4.1)] i.e.
$$ \alpha\,=\,\alpha(x), \quad \beta\,=\,\beta(x), \quad \gamma\,=\,
\gamma(x), \quad \mu\,=\,\mu(x);$$
It means in (4.1) all the components 
of the metrical tensor depend on the single spatial coordinate $x\, \in\, 
[x_0, \,x_a],$ where $x_a$ is the value of $x$ on the axis of symmetry, 
defined by the condition $\mbox{exp}[\beta(x_a)]\,=\,0$, and $x_0$ is the 
value of $x$ on the surface of the critical cylinder.  The coordinates $z$ 
and $\phi$ take their standard values: $z\,\in\,[-\infty,\,\infty], \quad 
\phi\,\in\, [0,2\pi].$  

(b) regularity of the metric and the matter fields in the whole 
space-time;

(c) localized in space-time (with limited field energy):
$$E_f\,=\,\int\,T_{0}^{0}\,\sqrt{-^3 g}\ dV \,<\, \infty.$$ 

Requirement (c) assumes the rapid decreasing of energy density of material 
field at spatial infinity, which together with (b) guaranties the 
finiteness of $E_f$. Let us note that $E_f$ may be finite even for
singular solutions on the axis. Requirement (b) means the regularity of 
material fields as well as the regularity of metric functions that entails 
the demand of finiteness of energy-momentum tensor of material fields all 
over the space. 

In view of requirement (a) it is convenient to choose the 
coordinate $x$ in (4.1) to satisfy the subsidiary condition 
[15]:  
$$ \alpha\,=\,\beta\,+\,\gamma\,+\,\mu, $$  
that permits to present the system of the Einstein equations  in the form: 
\begin{equation}
\mu^{\prime \prime}\,+\,\beta^{\prime \prime}\,-\, V\,=\,
-\,\kappa\,T_{0}^{0}\,e^{2 \alpha}, \end{equation} 
\begin{equation}
\mu^{\prime}\,\beta^{\prime}\,+\,\beta^{\prime}\,\gamma^{\prime}\,+
\,\gamma^{\prime}\,\mu^{\prime}\,=\,V\,=\,-\kappa\,T_{1}^{1}\,e^{2\alpha}, 
\end{equation} 
\begin{equation}
\gamma^{\prime \prime}\,+\,\beta^{\prime \prime}\,-\, V\,=\,
-\,\kappa\,T_{2}^{2}\,e^{2 \alpha}, \end{equation} 
\begin{equation}
\mu^{\prime \prime}\,+\,\gamma^{\prime \prime}\,-\, V\,=\,
-\,\kappa\,T_{3}^{3}\,e^{2 \alpha}, \end{equation} 

As in the preceding section, the electromagnetic 
field is described by the time component of the 4-potential ${\cal A}_0(x) 
\,=\,{\cal A}(x)$ and by the component ${\cal F}_{1\,0}\,=\,d{\cal 
A}/dx\,=\,{\cal A}^{\prime}$ of the field strength tensor and the 
energy-momentum tensor of interacting fields is defined by the 
equations (3.6), (3.7).  

Adding together the equations (4.3) and (4.4) and 
using  (3.7), one obtains the simple equation: 
\begin{equation}
\gamma^{\prime \prime}\,+\,\beta^{\prime \prime}\,=\,0,
\end{equation} 
with the solution 
\begin{equation}
\beta(x)\,+\,\gamma(x)\,=\,C_2\,x,\qquad C_2\,=\,\mbox{const.}
\end{equation}
Notice that the second 
integration constant in (4.7) can be taken trivial, as it determines only 
the choice of scale.  

In a similar way the addition of equations (4.3) and 
(4.5) leads to the equation:
\begin{equation}
\gamma^{\prime \prime}\,+\,\mu^{\prime \prime}\,=\,0,
\end{equation} 
with the solution 
\begin{equation}
\mu(x)\,+\,\gamma(x)\,=\,C_3\,x,\qquad C_3\,=\,\mbox{const.}
\end{equation}
Solving the equation (2.2) in the metric (4.1), 
one gets  the same result as in (3.9), i.e. 
\begin{equation}
\varphi^{\prime}(x)\,=\,C\,P(I).
\end{equation} 
Substituting (4.10) into (2.4),  one  finds the equation for the 
electromagnetic field, coincident with (3.10).  

Now as in the previous 
case, we use the equation (4.3) and  sum of equations (4.2) and (4.3) 
which in view of (4.6) and (4.8), take the form: 
\begin{equation}
\gamma^{\prime 2}\,-\,C_2\,C_3\,=\,-G\,\bigl(C^2\,P\,-\,{\cal A}^{\prime 
2}\,e^{-2\gamma} \bigr), 
\end{equation}
\begin{equation}
\gamma^{\prime \prime}\,=\,G\,e^{-2\gamma}\,\bigl({\cal A}^{\prime 2}\, 
+\,C^2\,{\cal A}^2\,P_I \bigr).
\end{equation} 
Elimination of $P_I\,{\cal A}$ between the 
equations (3.10) and (4.12) gives  the equation (3.13) with the solution 
(3.14). Integrating (3.14) under the choice $C_1\,=\,0,$ one obtains the 
equality (3.15). Finally, substituting $\gamma^{\prime}$ from (3.14) and 
$e^{2\gamma}$ from (3.15) into (4.11), one gets the equation for 
${\cal A}(x):$ 
\begin{equation} {\cal A}^{\prime 
2}\,(G\,{\cal A}^2\,+\,H)^{-2}\,=\, (G\,C^2\,P\,-\,C_2\,C_3)/\,G\,H. 
\end{equation}  
The equation (4.13) can be solved by quadrature:
\begin{equation}
\int\limits_{}^{}\frac{d {\cal A}}{(G\,{\cal A}^2\,+\,H)\, 
\sqrt{G\,C^2\,P\,-\,C_2\,C_3}}\,=\,\pm\,(1/\sqrt{G\,H})\, 
(x\,-\,x_0). 
\end{equation} 

Let us formulate regularity conditions to be satisfied by the 
solutions to the equations (2.2)-(2.4) on the  axis  of symmetry defined 
by the value $x\,=\,x_a,$ where $\beta(x_a)\,=\,-\infty.$  Choosing  in 
(4.7) $C_2\,<\,0$ and taking into account  that for the regular solutions 
$\mid \gamma \mid \,<\, \infty,$ we get $x_a \,=\, \infty.$ The regularity 
conditions are similar to (3.18) and (3.19) for the case of spherical 
symmetry, implying that the following relations hold as $x \to x_a \,=\, 
\infty:$
\begin{equation}
\gamma^{\prime}\,\to\,0, \quad {\cal A}^{\prime} \to 
{\cal A}_c\,\ne\,\infty, \quad {\cal A}^{\prime}\to 0, 
\end{equation} 
\begin{equation}
e^{2\mid C_2\mid x}\,P(I) \to 0, \quad 
e^{2\mid C_2\mid x}\,\mid I\,P_I \mid\,<\,\infty, \quad C_3\,\ne\,0.
\end{equation}
Boundary conditions on the surface of the critical cylinder $x\,=\, 
x_a$ can be written as follows:  
\begin{equation}
T_{\mu}^{\nu}\,=\,{\cal A}\,=\,{\cal A}^{\prime}\,=\,0, \quad e^{\gamma} 
\,=\,1, \quad e^{\beta}\,=\,e^{-\mid C_2 \mid x}\,>\,0.
\end{equation}
The conditions (4.17) together with the relations $e^{2\gamma}\,=\,G\, 
{\cal A}^2 \,+\,H,$ imply that $H=1.$  

Choosing $P(I)$ in the form (3.21), 
one can find the expression for ${\cal A}(x)$ which is similar to (3.22): 
\begin{equation} {\cal A}(x)\,=\,(1/\sqrt{G})\,\bigl[1\,-\, 
\mbox{exp}\,\bigl(-\frac{2\,C\,\sqrt{G}}{\sigma}\,(x\,-\,x_0)\bigr) 
\bigr]^{\sigma/2}.
\end{equation}  
As one can readily see from (4.18), 
the conditions (4.15), (4.16) and (4.17) are fulfilled if $\mid C_2 \mid 
\,\le\, C\,\sqrt{G}/\sigma.$  It is noteworthy that at $x\,\le\,x_0, \quad 
{\cal A}(x)\,\equiv\,0$ and the space-time is flat, the gravitational 
field being absent [17].  

There is a principal difference between  solutions 
(3.22) and (4.18).  For the case of spherical symmetry the droplet-like 
solution can be transformed to the soliton-like one if  the boundary 
$\xi_0$ is removed by putting $\xi_0\,=\,0$ (as in this case $\mbox{exp} 
[\beta(\xi_0)]\,=\,1/\xi_0\,=\,\infty$).  On the contrary, for the case 
of cylindrical symmetry the removal of the boundary is equivalent to 
putting $x_0\,=\,-\infty,$ as in this case $\mbox{exp}[\beta(x_0)] \,=\, 
\mbox{exp}(-\mid C_2 \mid x_0)\,=\, \infty.$  Under this last choice the 
solution (4.18) takes constant value ${\cal A}(x)\,=\,1/\sqrt{G}$ and  
the soliton structure disappears.  For the considered case, as  well as  
for that of spherical symmetry, the density of the field energy  is given 
by equation (3.25) and the  linear density of energy is similar to (3.27):
\begin{equation}
E_f\,=\,(C/4)\,\int\limits_{0}^{1/\sqrt{G}}\,d{\cal A}\,e^{-3\gamma}\, 
\bigl[\sqrt{P}(1\,+\,e^{2\gamma})\,+\,4\,I\,(\sqrt{P})_I\,\bigr], 
\end{equation} 
Substituting $P(I)$ from (3.21) into 
(4.19), one can find that $E_f$ is finite and the total energy 
$E_f\,+\,E_g$ turns out to be zero.  

Let us now define the effective charge density $\rho_e$ and total charge 
$Q$, corresponding to the unit length on z-axis. In generally from (2.4) 
one gets [16] \begin{equation} j^\alpha\,=\,\frac{1}{4\pi} \, 
\bigl(\varphi_{, \beta}\,{\varphi}^{, \beta}\bigr)\, \Psi_I \, {\cal 
A}^\alpha, \end{equation} that for static radial electric field leads to 
\begin{equation}
j^0\,=\,\frac{C^2}{4\pi}e^{-2(\alpha+\gamma)}\, P_I\,{\cal A}.
\end{equation}
Then for chronometric invariant electric charge density $\rho_e$ we have
\begin{equation}
\rho_e\,=\,\frac{j^0}{\sqrt{g^{00}}}\,=\,
\frac{C^2}{4\pi}e^{-(2\alpha+\gamma)}\, P_I\,{\cal A}.
\end{equation}
The total charge is defined from the equality
\begin{equation}
Q\,=\,2\pi\int\limits_{x_a}^{x_\infty}\rho_e\sqrt{-^3 g}\,dx.
\end{equation}
Putting the corresponding quantities  into the foregoing equality after 
some simple calculations we obtain
\begin{equation}
Q\,=\,\frac{1}{2}\,e^{-2\gamma}\,{\cal 
A}^{\prime}\mid_{x_a}^{x_\infty}\,=\,0.\end{equation} 

Now it is worthwhile to make again 
the comparison with the flat-space solutions of the equations (2.3) and 
(2.4), using  the interval:  
$$ ds^2\,=\, dt^2 \,-\,d\rho^2\,-\,\rho^2\, d\phi^2\,-\,dz^2.$$  
In this 
case the scalar field equation (2.3) admits the solution:
\begin{equation}
\varphi^{\prime}(\rho)\,=\,C\,P(I)/\rho, \quad P(I)\,=\,1/\Psi(I), \quad 
C\,=\,\mbox{const.}
\end{equation}
Inserting (4.25) into (2.4), one can 
find the electromagnetic field equation which admits the solution in 
quadratures: 
\begin{equation}
\int\limits_{}^{} \frac{d{\cal 
A}}{\sqrt{P(I)}}\,=\,\pm\,C\,\mbox{ln}\frac{\rho}{\rho_0},\quad
\rho_0\,=\,\mbox{const.} 
\end{equation}
Substituting $P(I)$
from (3.30) in (4.26), one gets  the solution of the droplet-like form:  
\begin{equation}
{\cal 
A}(\rho)\,=\,(1/\sqrt{\lambda})\,\bigl[1\,-\,\bigl(\frac{\rho}{\rho_0} 
\bigr)^{2\,C\,\sqrt{\lambda}/\sigma}\bigr]^{\sigma/2}.
\end{equation}
One concludes from (4.27) that ${\cal A}(\rho\,\ge\,\rho_0)\,\equiv\,0.$ 
It means  that the electric charge of the system is zero . For 
the solution (4.27) the regularity conditions both on the axis $\rho\,=\, 
0$ and on the surface of the critical cylinder $\rho\,=\,\rho_0$ are 
fulfilled if $C\,\sqrt{\lambda}\,\ge\,\sigma.$  It  is noteworthy that 
in the case of cylindrical symmetry, both in the flat space-time and with 
account  of the  proper gravitational field, there do not exist any 
soliton-like solutions, as for the choice $\rho_0\,=\,\infty$ the 
solution (4.27) degenerates into the constant: ${\cal 
A}(\rho)\,=\,1/\sqrt{\lambda}.$  

The linear density of the field energy 
in flat space-time  can be found from the expression similar to (3.32), 
and as well as  in the case of spherical symmetry, it is equal to zero:
$$E_f\,=\,\frac{C}{2}\,\sqrt{P\,I}\mid_{{\cal A}(\rho_0)}^{{\cal A}(0)}
\,=\,0,$$ 
as was expected. 

\section{Discussion} \setcounter{equation}{0}

Before summing up the results we need to say two words about the interaction 
Lagrangian. In 1951 J. Schwinger [18] applied the method to compute the 
effective coupling between a zero spin neutral meson and the electromagnetic
field. To write the interaction between scalar and electromagnetic fields 
we use some arbitrary functions of electromagnetic field. Thus our approach 
to generate an effective Lagrangian generalizes the one proposed by 
Schwinnger. The particular choice of $P(I)$ was made to obtain droplet-like
configurations. Using $P(I)$ that differs from (3.21) one can obtain quite 
different a solution [see e.g. 16]. Let us now sum up the results obtained.

\noindent
Exact regular statically spherical- and cylindrical-symmetrical solutions 
with sharp boundary (droplet-like solutions) to the equations of scalar 
nonlinear electrodynamics in General Relativity have been obtained. It is 
shown that outside the droplet gravitational and electromagnetic fields 
remain absent i.e. total energy and total charge of the configuration are 
zero. We underline once more the principal difference between the 
droplet-like solutions with spherical symmetry and those with cylindrical 
one. In the first  case there exists a possibility of continuous 
transformation of the droplet-like configuration into the solitonian one 
by transporting the sharp boundary to the infinity. As for the second 
case, there is no such a possibility, and the soliton-like configuration 
disappears when the boundary is smoothed tending to the infinity.  Further 
we intend to study the interaction processes of droplets with 
external electromagnetic and gravitational fields and also the scattering 
of photons and electrons on droplets.

\vskip 5mm
\noindent
{\bf Acknowledgements}

\noindent
B. Saha would like to thank Prof. S. Randjbar-Daemi and ICTP High Energy 
Section for hospitality. 


\newpage
\begin{thebibliography}{99}
\bibitem{1}
~J.\ Werle 1977 {\it  Phys.  Lett. B} {\bf 71} (2) 367  

\bibitem{2}~J.\ Werle 1980 {\it  Phys. Lett. B} {\bf 95} (3,4), 391 

\bibitem{3}~ J.\ Werle 1981 {\it Acta Physica Polonica} 
{\bf B 12} (6) 601

\bibitem{4}~J.\ Werle 1988 {\it Acta Physica Polonica} {\bf B 19} (3) 
203

\bibitem{5} 
~K.\ A.\ Bronnikov,\, Yu.\ P.\ Rybakov, and G.\ N.\ Shikin 1991 
{\it Izvestiya Vuzov, Physics} (5) 24  

\bibitem{6} ~
Yu.\ P.\ Rybakov,\, B.\ Saha, and G.\ N.\ Shikin  1994
{\it Problems of Statistical Physics and Field Theory, Moscow, PFU} 18

\bibitem{7} ~ 
K.\ A.\ Bronnikov, and G.\ N.\ Shikin 1985
{\it Proc. of Sir A.\ Eddington Centenary Symp. on the Relativity Theory, 
Singapore, WS} {\bf 2} 196


\bibitem{8} ~K.\ A.\ Bronnikov, and  M.\ A.\ Kovalchuk 1980
{\it  J. Phys. A: Math. Gen.} {\bf 3} 187


\bibitem{9} 
~K.\ A.\ Bronnikov,\, Yu.\ P.\ Rybakov, and G.\ N.\ Shikin 1993 
{\it  Comm. in Theor. Phys.} {\bf 2} 19 

\bibitem{10} ~ K.\ A.\ Bronnikov,\, V.\ N.\ Melnikov,\, 
G.\ N.\ Shikin, and K.\ P.\ Stanukovich 1979  
{\it  Ann. Phys.} {\bf 118} (1) 84 

\bibitem{11} ~
K.\ A.\ Bronnikov 1973
{\it Acta Physica Polonica} {\bf B 4} (2) 251

\bibitem{12} ~
Yu.\ P.\ Rybakov,\, B.\ Saha, and G.\ N.\ Shikin 1994
{\it  Communications in Theor. Phys.} {\bf 3} (1) 67 

\bibitem{13} ~ Yu.\ P.\ Rybakov,\, B.\ Saha, and G.\ N.\ Shikin 1992 
{\it GR 13, Cordoba, Argentina} 66 

\bibitem{14} ~K.\ A.\ Bronnikov  1979
{\it Prob. Theor. Grav. and Elem. Part. Moscow, Energoizdat} (10) 
37

\bibitem{15} ~G.\ N.\ Shikin 1984
{\it  Prob.  Theor.  Grav.  and Elem.  Part.  Moscow, 
Energoizdat} (14) 85  

\bibitem{16} ~G.\ N.\ Shikin 1995 {\it URSS, Moscow} 88  

\bibitem{17} ~Yu.\ P.\ Rybakov,\, B.\ Saha, and G.\ N.\ Shikin 1993 
{\it 8 Russian Gravitation Conf., RGA} 193 

\bibitem{18}~J.\ Schwinger 1951
{\it  Phys. Rev.} {\bf 82} (5) 664	
\end{thebibliography}
 \end{document}